\documentclass[dvips]{acta}
\usepackage{supertabular,lscape,epsfig}
\usepackage{amssymb}
\usepackage{amsmath}
\usepackage{txfonts}

\SetPages{0}{0}

\SetVol{00}{2012}

\begin{document}

\begin{Titlepage}
\Title{On the apsidal-motion of thirteen eclipsing binaries} \Author{Zasche, P.,}{Astronomical
Institute, Faculty of Mathematics and Physics, Charles University Prague,
   CZ-180 00 Praha 8, V Hole\v{s}ovi\v{c}k\'ach 2, Czech Republic\\
e-mail: zasche@sirrah.troja.mff.cuni.cz}

\Received{Month Day, Year}
\end{Titlepage}

\Abstract{Main aim of this paper is the first light curve and apsidal motion analysis of thirteen
eccentric eclipsing binaries and to reveal their basic physical properties. All of the systems were
studied by method of period analysis of times of minima and the light curve analysis. Many new
times of minima for all of the systems were derived and collected from the data obtained by the
automatic, robotic or satellite telescopes. This allow us to study the apsidal motion in these
systems in detail for the first time. From the light curve analysis the first rough estimations of
the physical properties of these systems were done. The analyzed systems undergo an apsidal motion
with the following periods in years: AR~CMa (44$\pm$10), OZ~Hya (117$\pm$53), V498~Mon (62$\pm$4),
V521~Mon (217$\pm$37), V684~Mon (74.5$\pm$20), V730~Mon (39$\pm$12), GV~Nor (197$\pm$67), NS~Nor
(516$\pm$230), TZ~Pyx (157$\pm$37), V385~Sco (1926$\pm$980), V629~Sco (56$\pm$17), V881~Sco
(131$\pm$48), V1082~Sco (186$\pm$280). Also 190 new minima times were derived and included. The
period and light curve analyses were done, however a more detailed spectroscopic analysis is needed
to confirm the physical parameters of the components with higher conclusiveness.}
 {binaries: eclipsing -- stars: fundamental parameters -- stars: individual: AR~CMa, OZ~Hya, V498~Mon,
  V521~Mon, V684~Mon, V730~Mon, GV~Nor, NS~Nor, TZ~Pyx, V385~Sco, V629~Sco, V881~Sco, V1082~ Sco}

\section{Introduction}

Testing the General Relativity as well as the stellar structure models of stars is possible thanks
to the eccentric eclipsing binaries. The investigation of period changes in these systems only on
the basis of times-of-minima variation (both primary and secondary ones) is a familiar method in
stellar astrophysics. It was described elsewhere, e.g. Gim{\'e}nez \& Garc{\'{\i}}a-Pelayo (1983),
or Gim{\'e}nez \& Bastero (1995). Only for a short recap of the method: the sidereal and
anomalistic periods of the binary are connected via a relation $P_s = P_a (1-\dot \omega /{2\pi}),$
where $\dot \omega$ is the rate of the apsidal advance $\omega = \omega_0 + \dot \omega E.$ The
period of such motion is $U = 2\pi P_a/\dot\omega.$ The individual equations for computing the time
of primary and secondary minima are given in Gim{\'e}nez \& Garc{\'{\i}}a-Pelayo (1983).

The number of new observations of the eclipsing binary (hereafter EB) systems is increasing every year
, but for some of the systems the detailed analysis is still missing due to the insufficient data coverage. 
For our analysis we used the data from the automatic photometric surveys - such as ASAS
(Pojmanski 2002), and NSVS (Wo{\'z}niak et al. 2004), 
as well as from the satellite data - the OMC camera onboard the INTEGRAL satellite, see Mas - Hesse
et al. (2004), and COROT\footnote{The CoRoT space mission was developed and is operated by the
French space agency CNES, with the participation of ESAs RSSD and Science Programmes, Austria,
Belgium, Brazil, Germany, and Spain. For online data see: http://idoc-corot.ias.u-psud.fr/}
satellite, see Baglin et al. (2006).

Despite rather low amplitudes of the light curves (none of the systems has the primary minimum as
deep as one magnitude in $V$~filter) the ASAS data obtained during the last decade were used for
the light curve analysis. The results of this analysis together with the apsidal motion study help
us to estimate the internal structure constants of the particular system.

An overview of the parameters for the systems presented in this paper, taken from already published
papers and/or databases (SIMBAD, GCVS - Samus et al. 2004) are given in Table 1, however some of
them are still only estimations, e.g. the $B-V$ index from the NOMAD survey (Zacharias et al. 2004)
or spectral types from Svechnikov \& Kuznetsova (1990). There is also listed the number of all
available times of minima nowadays (including our new ones) as well as the magnitude in $V$ filter
together with the depths of both minima for each system (taken from the ASAS survey). However, for
most of the systems the $B-V$ index from the photometry seems to be rather high (i.e. very reddened
stars), due to the fact that the eccentric systems are usually of early spectral type (e.g. in
catalogue of eccentric EBs by Bulut \& Demircan (2007) are 105 out of 124 systems of O, B, or A
spectral types.). Also for the systems where the spectral types are known, their respective $B-V$
indexes should be much lower. This indicates rather distant objects and that the interstellar
extinction was not subtracted.

\begin{table*}
 \footnotesize
 \caption{Basic information about the analyzed systems, taken from the literature.}
 \label{Table1} \centering \scalebox{0.85}{
\begin{tabular}{c c c c c c c c c c c}
\hline \hline
   Star   & Mag B & Mag V & (B-V) & (B-V) & Sp. &     Sp.    &  Min & Mag (V)& Mag  & Mag   \\
          &  G/S$^{\mathrm{a}}$  &  G/S$^{\mathrm{a}}$  &  G/S$^{\mathrm{a}}$  & NOMAD &     &    S\&K$^{\mathrm{b}}$   &      &  ASAS  & MinI & MinII \\
 \hline
   AR CMa & 11.9  &       &       & 0.263 &     & (F0)+[G8IV]&   14 & 10.88 & 11.65 & 11.59 \\
   OZ Hya &  9.9  &       &       &       & F2V &            &   11 &  9.45 &  9.62 &  9.61 \\
 V498 Mon & 10.49 &       &       & 0.322 & B5  &  (B9)+[F6] &   45 & 10.00 & 10.35 & 10.34 \\
 V521 Mon & 10.23 & 10.19 &  0.04 & 0.046 & B9  &   A0+[A1]  &   29 & 10.08 & 10.45 & 10.40 \\
 V684 Mon &  8.31 &  8.44 & -0.13 &-0.164 & B4  &            &   44 &  8.32 &  8.41 &  8.41 \\
 V730 Mon &  8.87 &  8.86 &  0.01 &       & B4  &            &   14 &  8.83 &  8.91 &  8.91 \\
   GV Nor & 10.66 & 10.6  &  0.06 & 0.102 &B5/8 & (A8)+[K0IV]&   25 & 10.55 & 11.08 & 11.05 \\
   NS Nor & 10.3  &       &       & 0.130 &A0IV &            &   16 & 10.11 & 10.62 & 10.43 \\
   TZ Pyx & 10.99 & 10.7  &  0.27 & 0.266 & A   & (A2)+[A2]  &   16 & 10.67 & 11.36 & 11.36 \\
 V385 Sco &       & 11.0  &       &-0.040 & B9  & B9+[G8IV]  &   10 & 10.93 & 11.47 & 11.42 \\
 V629 Sco & 11.9  &       &       & 0.570 &     & (A2)+[K0IV]&   13 & 11.63 & 12.04 & 12.01 \\
 V881 Sco &  9.61 &       &       & 0.364 &A1/A2IV& A0+[F0]  &   14 &  9.23 &  9.67 &  9.67 \\
V1082 Sco & 11.50 & 10.09 &  1.41 & 1.374 &B0.5Ia&           &   14 & 10.11 & 10.45 & 10.38 \\
\hline
\end{tabular} } \\
   {Note: $^a$ - G/S - taken from GCVS and/or SIMBAD} , {$^b$ - S\&K - Svechnikov \& Kuznetsova (1990)}
\end{table*}

\section{An approach for the analysis}

Because the spectroscopic study is missing for most of the systems in our sample, there are several
assumptions which have to be considered. For this reason we used a following approach for the
analysis:
 \begin{itemize}
   \item At the beginning all of the available minima were analyzed and a preliminary apsidal
   motion parameters were derived (with the assumption $i=90^\circ$).\\[-5mm]

   \item At second, the eccentricity ($e$), argument of periastron ($\omega$) and apsidal motion
   rate ($\dot \omega$) as resulted from the apsidal motion analysis were used for the preliminary
   light curve (hereafter LC) analysis.\\[-5mm]

   \item At third step the inclination ($i$) from the LC analysis was used for the apsidal motion
   analysis.\\[-5mm]

   \item And finally, the resulted $e$, $\omega$, and $\dot \omega$ values from the apsidal motion
   analysis were used for the final LC analysis.\\[-5mm]
 \end{itemize}

For the light-curve analysis we used the program {\sc PHOEBE} (Pr{\v s}a \& Zwitter 2005), which is
based on the Wilson-Devinney algorithm (Wilson \& Devinney 1971). For the systems where the radial
velocity (hereafter RV) study was performed the spectroscopic mass ratio $q_{sp}$ was used for the
LC analysis. Otherwise, after a few trials with the "q-search method" which resulted in $q_{ph} =
1.0$ for most cases, we chose the mass ratio equal to unity for all of the systems with unknown
$q_{sp}$. This is due to only negligible ellipsoidal variations outside of minima for most of the
systems and difficulties with a photometric mass ratio in detached binaries as quoted e.g. by
Terrell \& Wilson (2005). The temperature of the primary component was always fixed at a value of
typical temperature for a particular spectral type (e.g. Popper (1980), Harmanec (1988), or
Andersen (1991)). For a given mass ratio the semi-major axis was fixed at an appropriate value for
the primary mass to be equal to a typical mass of a particular spectral type. With this approach
besides the masses also the relative radii for both components can be roughly estimated for a
prospective derivation of the internal structure constant for the individual system (see below
Section 4). However, we would like to emphasize once more that these fundamental parameters are
still only rough estimates and have not been derived independently from calibrations, hence should
not be used as fundamental parameter sources.

\begin{figure*}
  \centering
  \includegraphics[width=1.0\textwidth]{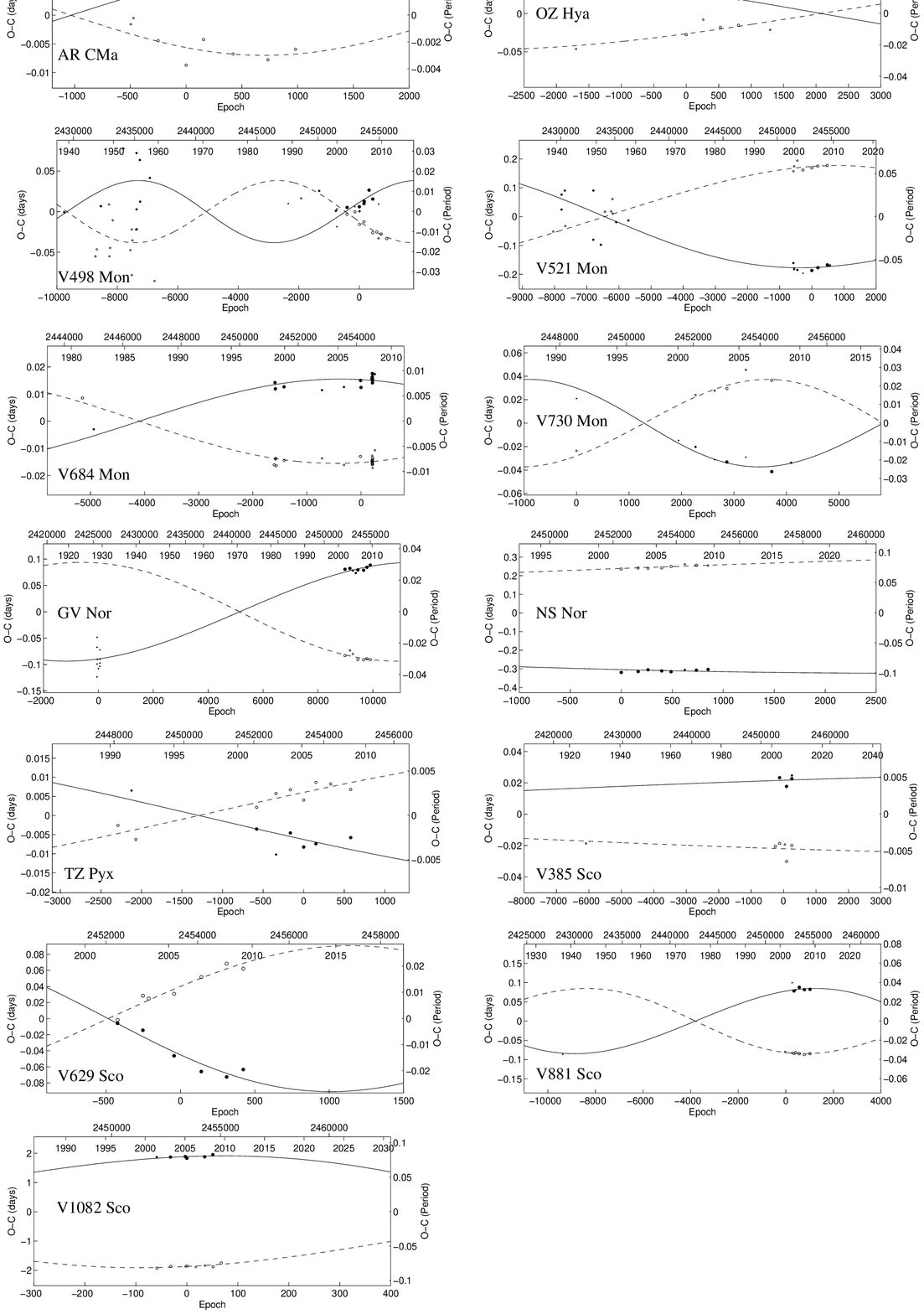}
  \FigCap{$O-C$ diagrams of the studied systems. The lines represent the fit according to the apsidal
  motion hypothesis (see text and Table 2), solid line for the primary, while the dashed
  line for the secondary minima, dots stand for the primary and open circles for the secondary minima,
  bigger the symbol, higher the weight.}
  \label{FigOC}
\end{figure*}

\section{The individual systems}

\subsection{AR CMa}

AR~CMa was discovered as an eclipsing variable in 1934, van Hoof (1943). The authors mentioned its
orbital period of 1.16607~day and no secondary minimum. Since then this orbital period was adopted
as a correct one and no other detailed study of this system was published.

We used the ASAS data, which cover the time epoch from 2001 to 2009. The whole light curve of AR
CMa was observed in $V$ filter, and these data were used for deriving several new times of primary
and secondary minima (see the Table 5 available in its entirety in the electronic version of the
journal). The Kwee \& van Woerden (1956) method was used for the minima computation. We found that
the orbital period is double the value presented by van Hoof (1943), while the minima are of 0.77
and 0.71~mag deep, respectively (see Table 1).

The $O-C$ diagram of all available minima as derived from the ASAS, Rotse, and NSVS surveys is
shown in Fig.1. We followed the method of apsidal motion analysis as described in Gim{\'e}nez \&
Garc{\'{\i}}a-Pelayo (1983). In Fig. 1 we plot the final fit with the apsidal motion hypothesis.
The parameters of such fit are given in Table 2.

The LC analysis resulted in fit presented in Fig. 2 and the final LC parameters are given in Table
3. This fit was derived assuming the primary component of A9 spectral type (as derived from the
$B-V$ indexes from NOMAD, Tycho, Kharchenko (2001), etc.), therefore the temperature of the primary
component $T_1=7250$~K was fixed. The derived value of internal structure constant is given below
in section 4.

\begin{figure*}
  \centering
  \includegraphics[width=1.0\textwidth]{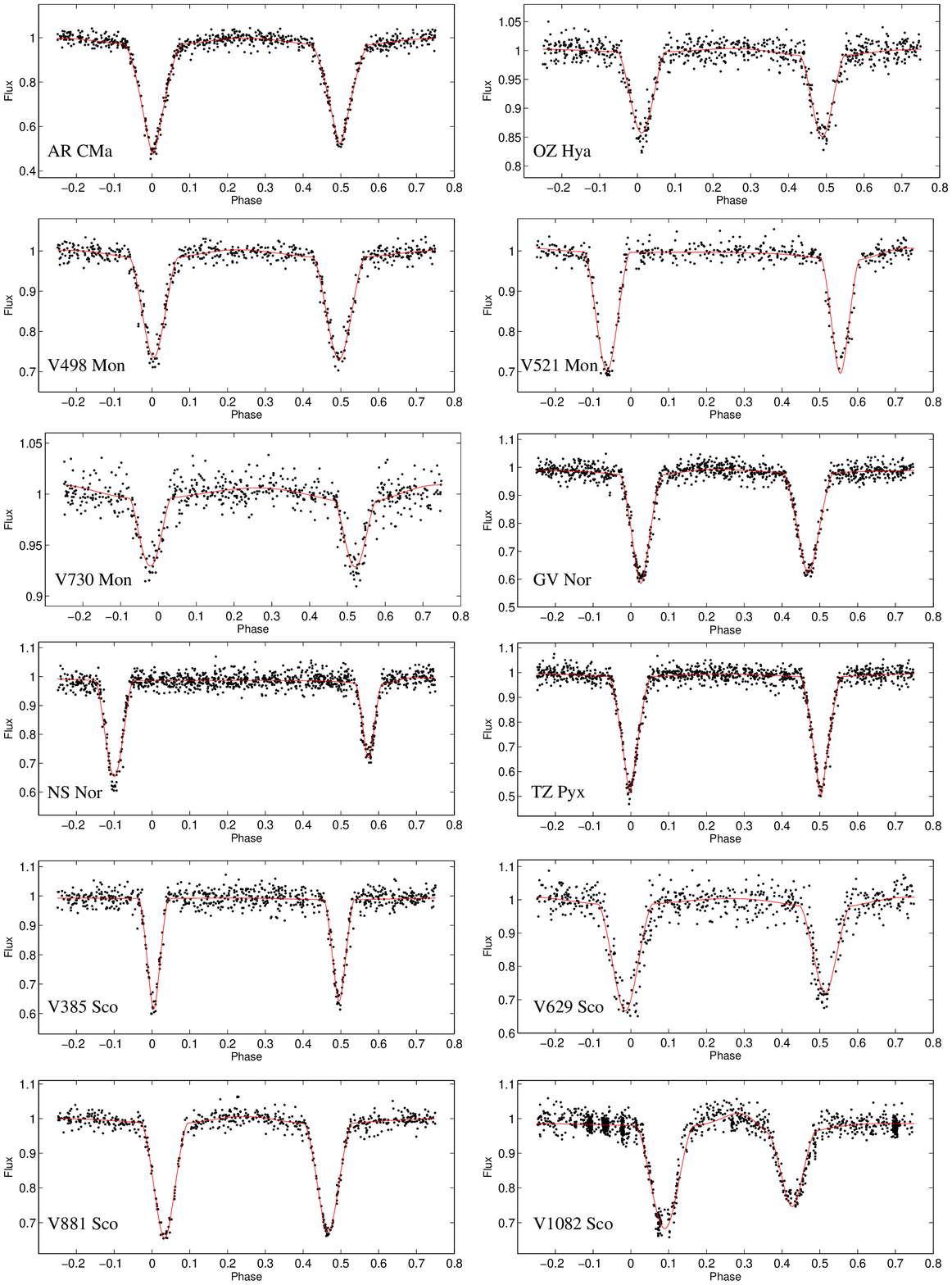}
  \FigCap{The light curves of the analyzed systems. The individual points represent the ASAS data,
  while the lines stand for the final fits according to the parameters presented in Table 3.}
  \label{FigsLC}
\end{figure*}

\subsection{OZ Hya}

OZ Hya (= HIP 49177) was discovered as a variable star by Hoffmeister (1936). It is a member of
visual double designated as B~871, or WDS J10022-1946AB in the Washington Double
Star Catalog (WDS
, Mason et~al. 2001). There were obtained 15 astrometric measurements since its discovery as a
double in 1927, but these data indicate no visible motion.

The spectral type of the system is listed as F2V, according to Houk \& Smith-Moore (1988). We can
only speculate which component this spectral type belongs to - whether the visual component or the
EB pair. It was also included into the catalogue of systems located in the $\delta$~Sct region of
the Cepheid instability strip (Soydugan et al. 2006), which are suspect to contain pulsating
component. However, the quality of the ASAS data used in our analysis cannot prove or rule out this
hypothesis.

Analysing all minima times (Table 5) we derived very preliminary apsidal motion parameters (Table 2
and Fig. 1). The analysis of LC yielded Fig. 2 and the parameters given in Table 3 from which the
most interesting seems to be the value of the third light, which resulted in about 40\% of the
total luminosity of the system. This is in excellent agreement with the fact that the magnitude
difference between the close visual component and the eclipsing binary itself is about 0.65~mag in
V filter (WDS).

\subsection{V498 Mon}

V498 Mon (= HD 288904) was discovered as a variable star by Wachmann (1966). Its spectral type was
classified as B5 by Nesterov et al. (1995). For the analysis we collected old photographic minima
together with new ones from NSVS and ASAS. The most recent minimum was observed in the Ond\v{r}ejov
observatory with 65-cm telescope equipped with the CCD camera.

As one can see from the fit in Fig. 1, the motion of line of apsides is relatively fast, with
period about 62~years only and is evident even in the ASAS data covering about 10~years. The
difference between primary and secondary minima is still increasing, reaching its maximum in the
year 2017. The LC analysis was carried out with the assumption of the primary component's
temperature $T_1=15500$~K (according to B5 spectral type).

\subsection{V521 Mon}

V521 Mon (= HD 292704) is an Algol-type EB, discovered also by Wachmann (1966). McCuskey (1956)
presented its spectral type as A0, while Brancewicz \& Dworak (1980) reported B7 spectral type, and
Nesterov et al. (1995) lists spectrum of B9 type. On the other hand, the $UBV$ and $uvby\beta$
photometry by Lacy (1992) and Lacy (2002) indicates a bit later spectral type, about A5-A7
(according to Popper 1980). 

The analysis of all available minima led to the parameters given in Table 2. The apsidal motion
period is rather long, about 217~years, but about 1/3 is already covered by observations. After the
subtraction of the apsidal motion term, the residuals show some additional variation, which can be
caused e.g. by a third body in the system. However, it still remains rather uncertain now and
further observations are needed. For the LC analysis the primary component's temperature was fixed
at a value of 10350~K (sp B9). No additional third light was detected.

\subsection{V684 Mon}

V684 Mon (= HD 47755) was discovered by Koch et al. (1986). It is also a member of very young open
cluster NGC 2264 (cluster age is 3-4 Myr), as well as primary of a visual double star WDS
J06406+0947 (where the secondary component about 25\arcs~ distant is also variable V780~Mon). A
detailed analysis of LC and RV curves was performed by Bradstreet et al. (2007). There resulted
that the system consists of two main sequence B4 stars on slightly eccentric orbit.

For our analysis we used the minima from the abovementioned analysis (Bradstreet 2011, private
communication), together with the ASAS and COROT data. The scatter of individual minima is caused
by shallow eclipses and only poor photometric data for minima computation. However, our
eccentricity is very close to that one published by Bradstreet et al. (2007). About 1/3 of the
apsidal period is already covered by observations.

\subsection{V730 Mon}

V730 Mon (= HD 46738 = HIP 31371) is the system with the shortest period in our sample. It is a
primary component of a visual pair A~508 (or WDS J06347-0836AB), which was discovered as a double
in 1903, but since then there was no mutual motion of the two components detected. Kharchenko
(2001) published the only spectral classification of both visual components as B3V (component A:
V730~Mon) + B4 (fainter component B).

Analysing the new derived minima, there arises that the period of apsidal motion is about 39~yr
only and about 1/3 is covered by observations now. The LC analysis was carried out with the
assumption of spectral types by Kharchenko (2001), i.e. $T_1=19000$~K. As one can see from the
individual luminosities, the triple system is probably consisted of three very similar stars.

\subsection{GV Nor}

GV Nor (= HD 146375) was discovered as a variable by Kruytbosch (1932). The only spectral
classification of the star is that by Houk \& Cowley (1975), who gave its spectral type B5/8,
unfortunately based on poor quality spectra.

Our period analysis of GV~Nor is based on 25 minima in total, resulting in apsidal period of about
200~years. After subtracting the fit, some additional variation of the minima cannot be ruled out.
The LC analysis was performed under assumption that the primary component is of B7 spectral type
($T_1=13000$~K). The LC fit shows clearly defined and relatively deep eclipses, which are similar
to each other.

\subsection{NS Nor}

NS Nor (= HD 147944) was discovered as a variable star by Hoffmeister (1963), after then Meinunger
(1970) listed its type as a long-periodic with period about 130~days. SIMBAD lists NS Nor as a
"Semi-regular pulsating Star". However, no such behavior was found in the ASAS data (also
INTEGRAL/OMC and "Pi of the sky" data show no such variations). Its spectral type was classified as
A0IV by Houk \& Cowley (1975).

Our apsidal motion analysis is still only very preliminary yet due to its long period. The LC
analysis was carried out with the assumption of primary temperature $T_1=9400$~K. However, we can
still doubt about the initial assumption of the analysis, which is the spectral classification as a
subgiant, because a more evolved star should probably be older and its orbit definitely
circularized. Therefore, a detailed spectroscopic analysis is still needed.

\subsection{TZ Pyx}

TZ Pyx (= HIP 42619) is probably the most studied system in our sample. It was discovered as a
variable one by Strohmeier (1966). The spectroscopic analysis by Duerbeck \& Rucinski (2007)
revealed that the system is a well-detached one and its spectral type was classified as A with both
components very similar to each other (BFs, mass ratio, and depths of both minima). However, their
RV study deals with one disadvantage, which is the circular orbit assumed for the analysis. Otero
(2007) analysed the Hipparcos and ASAS data and found out that the system is definitely an
eccentric one.

Our analysis is based on the observations after 1990's only, because the older published minima
have much larger scatter than necessary for the apsidal motion analysis performed here. The apsidal
motion fit is still not very convincing and would have even shorter period (the two most recent
deviating points), but only further observations would confirm this hypothesis.
 For the LC analysis we used the assumption that the primary component is of A1 spectral
type. This is based on the $B-V$ indices from different sources as well as from Ammons et al.
(2006). Hence, we set the value of $T_1=9100$~K. Both eclipses are relatively well-covered by the
data points and also the fit is satisfactory. The mass ratio $q$ was taken from the RV analysis by
Duerbeck \& Rucinski (2007).

\subsection{V385 Sco}

The star V385 Sco (= HD 320236) was discovered as an eclipsing variable by Swope (1940), who also
gave its orbital period of 2.34515~day. Its true orbital period is double, about 4.69~day. It is
probably a member of open cluster Collinder 338. No detailed analysis of this system was performed,
only the spectral type was estimated as B9 by Parsons et al. (1980).

Using all available minima we performed an apsidal motion analysis. Unfortunately, the apsidal
motion is still very poorly covered by the data, however the apsidal advance is evident thanks to
the one old data point. New observations are still needed. The LC analysis was done assuming the
primary temperature $T_1=10350$~K.

\subsection{V629 Sco}

The system V629 Sco was discovered by Swope (1943), who also wrote a remark about its apsidal
motion, which means that the primary and secondary minima were observed both. Unfortunately, only
time of primary minimum is listed there. There is also a remark that "Further details will be
published elsewhere," but no other publication about this concern was found in the literature.
Svechnikov \& Kuznetsova (1990) presented the spectral type (A2)+[K0IV], which is quite unreliable
for the secondary, but would be more likely for the primary component.

New minima cover only a few years but even on these data there is evident a rapid apsidal motion,
which has the period about 56~years only. Further observations are still needed for the parameters
to be derived with higher conclusiveness. The LC analysis was carried out with the initial
assumption of $T_1=8770$~K. As one can see from the fit plotted in Fig. 2, the apsidal motion is so
rapid that it even cause a "blurring" of a phase light curve (during the time epoch about 8.7~years
of ASAS data the argument of periastron $\omega$ changed of about 56$^\circ$). The scatter of the
light curve is rather high and new more precise observations would be very welcome.

\subsection{V881 Sco}

V881 Sco (= HD 150384) was discovered as a variable by Strohmeier \& Knigge (1973), and Houk (1982)
gives its spectral classification as A1/2IV.

The apsidal motion analysis was performed and thanks to the one old minimum from 1930's there is
about one half of the apsidal motion period covered now. However, due to lack of data, the analysis
is still only preliminary yet. The LC solution was done with the assumption of $T_1=9100$~K
(spectral type A1 for the primary).

\subsection{V1082 Sco}

V1082 Sco (= HD 318015 = HIP 86163) is a member of very reddened young open cluster Trumpler 27,
see The \& Stokes (1970) and Moffat et al. (1977). Its spectrum was classified as B0.5Ia by
Drilling \& Perry (1981).

Available minima times show very slow apsidal motion, but only a small part of its period is
covered yet and further observations are needed. The LC of rather interesting shape is plotted in
Fig. 2, however there is still a place for doubts if the eclipses are total or not. Regrettably,
the ASAS photometry is not precise enough for a final confirmation of this hypothesis. Secondary
minimum is located near the phase $\phi_{II} = 0.33$ with respect to the primary one, which
indicates rather high value of eccentricity. This is the system which shows the largest
outside-eclipse variations in our sample. We found a light curve solution (presented in Table 3)
using the assumption of $T_1=28500$~K.

\begin{table*}
 \caption{The parameters of the apsidal motion fits.}
 \label{OCparam}
 \scriptsize
 \centering \scalebox{0.81}{
 \begin{tabular}{c c c c c c c c c}
 \hline\hline
  System  & $HJD_0 - 2400000$ &  $P$ [d]   &   $P_a$ [d]   &    $e$     & $\omega$ [deg] & $\dot \omega$ [deg/cycle] & $U$ [yr] & $\Sigma$ res$^2$ \\
 \hline
 AR CMa   & 52648.8727(32) & 2.3322242(12) & 2.3325597(12) & 0.0094(33) &    143(12)     &      0.052(15)            & 44(10)   & 0.00026 \\
 OZ Hya   & 51983.9437(72) & 2.0487656(93) & 2.0488637(93) & 0.071(25)  &    234(27)     &      0.017(14)            & 117(53)  & 0.00267 \\
 V498 Mon & 53374.8246(60) & 2.4727687(14) & 2.4730374(14) & 0.048(7)   &    108(4)      &      0.039(3)             & 62(4)    & 0.10120 \\
 V521 Mon & 53203.035(28)  & 2.9706724(76) & 2.9707838(76) & 0.186(36)  &    357(5)      &      0.013(3)             & 217(37)  & 0.16876 \\
 V684 Mon & 54153.5363(24) & 1.8514196(17) & 1.8515456(17) & 0.025(6)   &    190(8)      &      0.024(9)             & 74(20)   & 0.00085 \\
 V730 Mon & 48502.744(21)  & 1.5723188(74) & 1.5724916(74) & 0.071(35)  &    219(21)     &      0.040(18)            & 39(12)   & 0.00123 \\
 GV Nor   & 25560.147(17)  & 2.9718631(38) & 2.9719861(38) & 0.098(23)  &    13(9)       &      0.015(8)             & 197(67)  & 0.00634 \\
 NS Nor   & 52258.49(13)   & 3.223580(124) & 3.223636(12)  & 0.308(162) &    329(45)     &      0.006(5)             & 516(230) & 0.00589 \\
 TZ Pyx   & 53423.1493(19) & 2.3185523(18) & 2.3186463(18) & 0.026(9)   &    289(10)     &      0.015(4)             & 157(37)  & 0.00061 \\
 V385 Sco & 53268.248(81)  & 4.6902421(42) & 4.6902734(42) & 0.018(12)  &    143(58)     &      0.002(2)             & 1926(980)& 0.00111 \\
 V629 Sco & 53626.932(12)  & 3.2491152(41) & 3.2496349(41) & 0.087(31)  &    298(13)     &      0.057(26)            & 56(17)   & 0.00582 \\
 V881 Sco & 52128.414(14)  & 2.4915700(66) & 2.4916993(66) & 0.106(17)  &    161(30)     &      0.019(11)            & 131(48)  & 0.00091 \\
V1082 Sco & 53454.32(54)   &23.4456(28)    &23.4537(28)    & 0.247(101) &    181(63)     &      0.124(74)            & 186(280) & 0.14563 \\
 \hline
 \end{tabular}}
\end{table*}

\begin{table*}
 \caption{The parameters of the light curve fits (values of temperatures are only estimations).}
 \label{LCparam}
 \tiny
 \centering \scalebox{0.8}{
 \begin{tabular}{c c c c c c c c c c c c c c}
 \hline\hline
  System  &   $i$     & $\Omega_1$& $\Omega_2$& $T_1$ &  $T_2$    &    $L_1$    &    $L_2$    &   $L_3$    &   $F_1$  &   $F_2$  & $R_1/a$  & $R_2/a$  & $\Sigma$ res$^2$ \\
          &   [deg]   &           &           & [K]   &   [K]     &    [\%]     &     [\%]    &    [\% ]   &          &          &          &          & \\
 \hline
 AR CMa   & 91.33(22) & 5.47(7)   &  5.30(6)  & 7250  &  7230(34) & 50.29(1.06) & 49.71(0.89) & 0.00(0.42) & 1.22(54) & 1.39(32) & 0.227(6) & 0.237(5) & 0.13778 \\
 OZ Hya   & 79.66(36) & 5.94(10)  &  5.42(9)  & 6700  &  7740(45) & 30.07(0.94) & 29.92(0.82) &40.01(1.33) & 1.14(65) & 1.03(50) & 0.202(9) & 0.211(7) & 0.06870 \\
 V498 Mon & 79.87(20) & 5.55(5)   &  5.81(6)  & 15500 &16255(141) & 49.92(0.98) & 50.08(0.99) & 0.00(0.84) & 1.15(21) & 1.19(25) & 0.224(4) & 0.211(4) & 0.06210 \\
 V521 Mon & 82.50(28) & 6.98(16)  &  6.55(9)  & 10350 &13343(132) & 49.80(1.30) & 50.20(1.22) & 0.00(1.07) & 1.27(51) & 1.34(45) & 0.176(8) & 0.190(5) & 0.05643 \\
 V730 Mon & 72.51(39) & 5.92(10)  &  5.42(7)  & 19000 &19594(287) & 33.32(1.01) & 33.28(0.90) &33.40(3.34) & 0.90(76) & 1.40(39) & 0.207(8) & 0.234(7) & 0.03882 \\
 GV Nor   & 85.54(22) & 6.42(7)   &  6.36(6)  & 13000 & 11382(88) & 50.07(0.65) & 49.93(0.78) & 0.00(0.65) & 1.21(23) & 1.12(20) & 0.191(3) & 0.193(4) & 0.19204 \\
 NS Nor   & 84.24(27) & 9.46(11)  &  7.79(11) & 9400  & 10789(81) & 50.86(1.02) & 49.14(1.01) & 0.00(0.99) & 0.77(69) & 1.23(53) & 0.125(3) & 0.159(5) & 0.24938 \\
 TZ Pyx   & 89.21(22) & 6.84(9)   &  6.74(8)  & 9100  &  9102(58) & 49.97(0.93) & 49.96(1.04) & 0.06(0.44) & 1.37(24) & 1.09(30) & 0.172(5) & 0.169(5) & 0.22866 \\
 V385 Sco & 86.68(40) & 9.11(24)  &  8.79(29) & 10350 &11953(121) & 49.73(1.37) & 50.27(1.29) & 0.00(1.24) & 1.44(83) & 1.20(71) & 0.116(7) & 0.121(8) & 0.16524 \\
 V629 Sco & 81.24(37) & 5.91(9)   &  5.33(7)  & 8770  &  8926(93) & 49.83(1.65) & 50.17(1.73) & 0.00(1.65) & 1.17(38) & 1.21(59) & 0.208(7) & 0.238(7) & 0.32045 \\
 V881 Sco & 83.32(20) & 6.22(7)   &  6.11(5)  & 9100  &  9249(63) & 49.73(0.91) & 50.27(0.94) & 0.00(0.71) & 1.20(30) & 1.20(45) & 0.197(5) & 0.202(4) & 0.06581 \\
V1082 Sco & 79.01(14) & 6.53(8)   &  5.33(6)  & 28500 &33322(242) & 49.81(0.87) & 50.19(0.84) & 0.00(0.63) & 1.71(43) & 1.25(18) & 0.198(5) & 0.259(2) & 0.22720 \\
 \hline
 \end{tabular}}
\end{table*}

\begin{table*}
 \caption{The internal structure constants $\log k_{2,obs}$ as compared with the theoretical
 values from stellar evolution models together with the relativistic fraction of the total apsidal
 motion rate.}
 \label{logK2}
 \small
 \centering \scalebox{0.8}{
 \begin{tabular}{c c c c || c c c c}
 \hline\hline
  System & log $k_{2,obs}$ & log $k_{2,theor}$ & $\omega_{rel}/\omega_{total}$ &   System & log $k_{2,obs}$ & log $k_{2,theor}$ & $\omega_{rel}/\omega_{total}$  \\
 \hline
 AR CMa   & -2.24 $\pm$ 0.20 & -2.39 $\pm$ 0.05 &  1.3 \% & NS Nor   & -2.45 $\pm$ 0.20 & -2.30 $\pm$ 0.05 & 12.2 \% \\
 OZ Hya   & -2.47 $\pm$ 0.20 & -2.30 $\pm$ 0.05 &  3.9 \% & TZ Pyx   & -2.14 $\pm$ 0.20 & -2.36 $\pm$ 0.05 &  5.5 \% \\
 V498 Mon & -2.23 $\pm$ 0.20 & -2.20 $\pm$ 0.05 &  3.2 \% & V385 Sco & -2.23 $\pm$ 0.20 & -2.35 $\pm$ 0.05 & 23.8 \% \\
 V521 Mon & -2.43 $\pm$ 0.20 & -2.35 $\pm$ 0.05 &  5.9 \% & V629 Sco & -2.13 $\pm$ 0.20 & -2.34 $\pm$ 0.05 &  1.1 \% \\
 V684 Mon & -2.10 $\pm$ 0.05 & -2.12 $\pm$ 0.05 &  6.9 \% & V881 Sco & -2.39 $\pm$ 0.20 & -2.34 $\pm$ 0.05 &  4.2 \% \\
 V730 Mon & -2.29 $\pm$ 0.20 & -2.15 $\pm$ 0.05 &  5.4 \% &V1082 Sco & -2.04 $\pm$ 0.20 & -1.91 $\pm$ 0.05 &  0.5 \% \\
 GV Nor   & -2.40 $\pm$ 0.20 & -2.23 $\pm$ 0.05 &  6.4 \% &  & & & \\
 \hline
 \end{tabular}}
\end{table*}

\section{Internal structure constant} \label{INT}

Thanks to the analysis of apsidal motion via the period changes and the light curve one is able to
use the derived quantities for a rough estimation of the internal structure constants. The
theoretical $\log k_{2,theo}$ values are taken from Claret (2004) and the comparison with the
observed ones is presented in Table 4. There are only mean values of $\log k_2$ of both components
which can be derived and these are based on several assumptions. The errors were only estimated due
to the fact that also the input parameters like spectral types are not known with high reliability.

The most speculative is the use of apsidal motion parameters for those systems, where only a short
time interval (compared with the apsidal motion period) is covered with the data nowadays. For
systems like OZ~Hya, NS~Nor, V385~Sco and V1082~Sco these values are still rather speculative and
affected by relatively large errors because of their poor coverage of minima times. Therefore,
almost the same chi square can be reached with slightly different values of apsidal motion rate and
hence also the estimation of $\log k_2$ is still only preliminary yet. Another possible sources of
errors for the internal structure estimation are the unknown spectral types of the components,
hence only rough estimation of their masses. For some systems also the interstellar reddening could
play a role when the spectral type was only estimated from the $B-V$ photometric index. And finally
of course the unknown value of the semimajor axis of the system (dealing with no RVs). Also better
the quality of the light curves, better the result of the LC fit and the parameters derived from
this fit. All of these are the possible sources of errors which affect the resulted $\log k_2$
values. However, for all systems the observed internal structure constants and the theoretical
values lie within their respective error bars (Table 4 and Fig. 3).

A direct comparison between theory and observations is still rather difficult here. As one can see
from Fig. 3, the distribution of resulting $\log k_{2,obs}$ values follow the predicted relation
(the earlier the spectral type yields the value of $\log k_2$ closer to zero) and all thirteen
values are randomly distributed around the theoretical ones. However, the error bars are still
large to do any definite finding concerning the evolutionary models. Generally we can say that
obtaining the precise RV curves for all of these systems would be of great benefit.


\begin{table*}
 \caption{Heliocentric minima used for the analysis.}
 \label{Minima}
 \footnotesize  
 \centering
 \begin{tabular}{l l l r r c c l}
 \hline\hline
 System & HJD - 2400000 & Error   &  Epoch  &   $O-C$  & Type & Filter & Reference\\
 \hline
 AR CMa   & 51493.2540  & 0.05    & -495.50 & -0.00158 & Sec & V  & Rotse \\
 AR CMa   & 51549.22847 & 0.035   & -471.50 & -0.00049 & Sec & -- & NSVS  \\
 AR CMa   & 52063.49034 & 0.04664 & -251.00 &  0.00593 & Pri & V  & ASAS  \\
 AR CMa   & 52064.64609 & 0.00625 & -250.50 & -0.00443 & Sec & V  & ASAS  \\
 AR CMa   & 52648.87782 & 0.01952 &    0.00 &  0.00513 & Pri & V  & ASAS  \\
 AR CMa   & 52650.03011 & 0.00322 &    0.50 & -0.00870 & Sec & V  & ASAS  \\
 \hline \hline
\end{tabular}
 \begin{list}{}{}
 \item Note: Table is published in its entirety in the electronic supplement of the journal.
 A portion is shown here for guidance regarding its form and content.
 \end{list}
\end{table*}

\section{Discussion and conclusions}

We derived the preliminary apsidal motion and light curve parameters for thirteen Algol-type
systems. None of these systems was studied by means of an apsidal motion hypothesis based on the
minima times analysis, and the LC solution is presented here for the first time. However, the
analysis still deals with several problems. The main disadvantage is that we have only very limited
information about these systems and our results are based only on rough estimations. The parameters
like spectral types of the components, their physical parameters (mass ratio) or the age of the
system are not known nowadays (except for analyses of TZ Pyx and V684~Mon) and further more
detailed study is very needed. Especially, the spectroscopic observations would be very welcome to
confirm the spectral types of both components, because the rough estimations based only on the
photometric indices are affected by relatively large errors (typical example is V521~Mon, where the
spectral estimations ranging between B7 and A7). Moreover, we found that the spectral estimation
published by Svechnikov \& Kuznetsova (1990) seems to be useless in most cases.

\begin{figure}
  \centering
  \includegraphics[width=70mm]{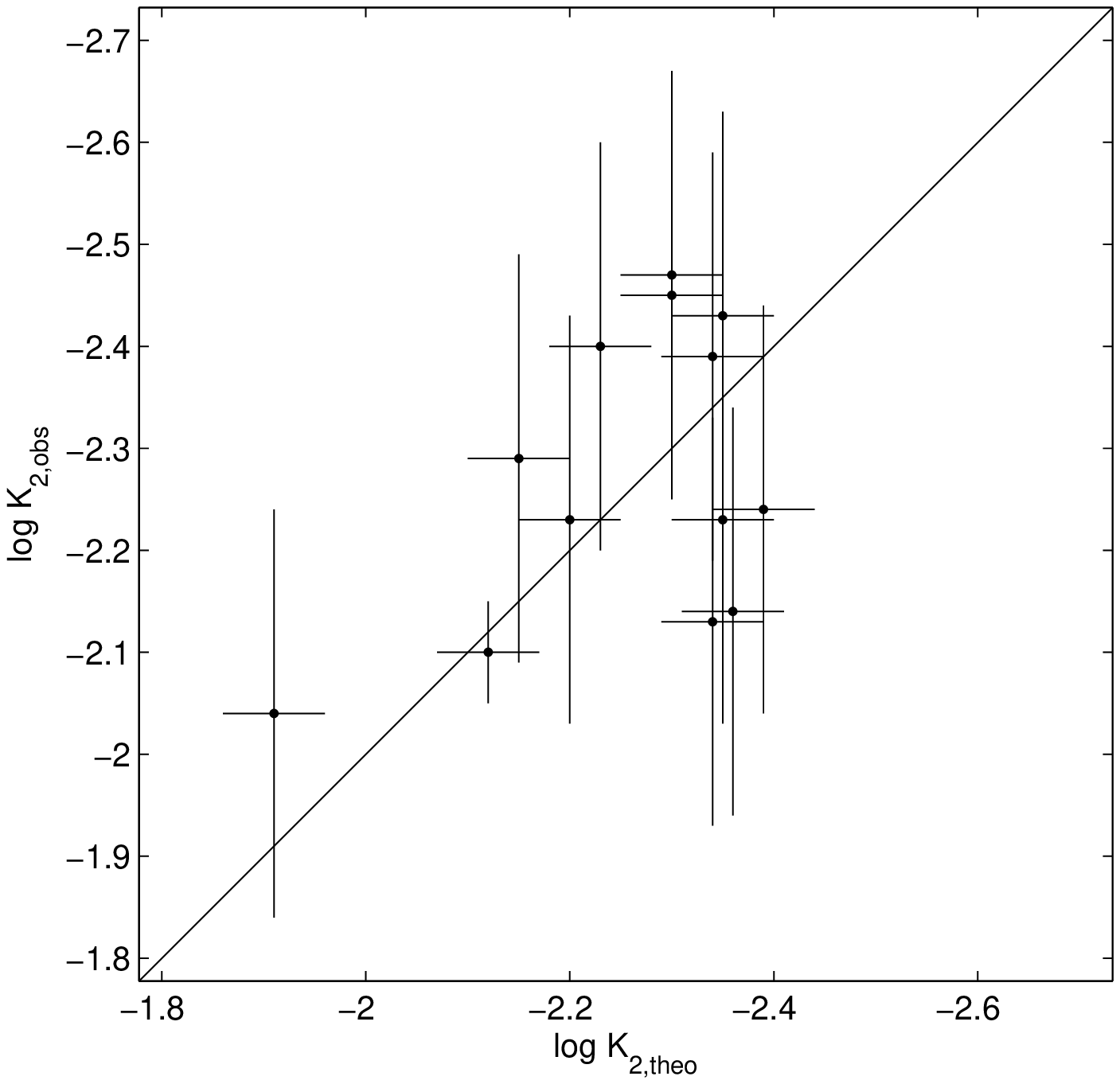}
  \FigCap{Plot of internal structure constants - observed versus theoretical.}
  \label{FigK2}
\end{figure}

The more detailed spectroscopic study for all of the systems would be also of great benefit for the
comparison of the internal structure constants and the models. Our presented first attempt on this
issue was done only on very preliminary physical parameters of the individual components and
therefore its findings on the resulting $\log k_2$ constants is also affected by large errors. Also
the new minima observations would be of great benefit for some of the systems, especially for those
of shorter apsidal periods (such as AR~CMa, V730~Mon or V629~Sco) and also for those, where we
noticed some suspected additional variations potentially caused by the third components in these
systems (such as V521~Mon or GV~Nor). Some of the stars selected for this sample are also possible
to observe from the northern hemisphere and the observations could be done rather easily
(relatively high brightness, moderately deep minima and not so long orbital periods). Most of the
data used for this study came from the ASAS survey, however the quality of these data is not
sufficient enough for a more detailed light curve analysis in most of these systems. New more
precise photometric observations in different photometric filters would be also welcome for a
prospective detailed LC analysis. New times of minima observations, whole light curves, as well as
a detailed spectroscopic analysis are still needed.

\Acknow{Based on data from the OMC Archive at LAEFF, pre-processed by ISDC. We thank the "ASAS",
"Pi of the sky", "NSVS", and "COROT" teams for making all of the observations easily public
available. This work was supported by the Czech Science Foundation grant no. P209/10/0715 and also
by the Research Programme MSM0021620860 of the Czech Ministry of Education. This research has made
use of the SIMBAD database, operated at CDS, Strasbourg, France, and of NASA's Astrophysics Data
System Bibliographic Services.}

\end{document}